\documentclass[superscriptaddress, 
reprint,
 amsmath,amssymb,
  aps, pre
]{revtex4-1}

\usepackage{newfloat}
\usepackage[noend]{algpseudocode}
\usepackage{mathptmx}
\usepackage{amsmath,amssymb}
\usepackage{graphicx}
\usepackage{times}
\usepackage{latexsym,float,epsfig,subcaption,footnote}
\usepackage{color}

\graphicspath{{./}}

\DeclareFloatingEnvironment[
    fileext=loa,
    listname=List of Algorithms,
    name=ALGORITHM,
    placement=tbhp,
]{algorithm}

\newcommand{\bigO}{\operatorname{O}}

\begin{document}


\title{Detecting vortices in superconductors: Extracting one-dimensional topological singularities from a discretized complex scalar field}


\author{Carolyn L. Phillips}
\email{corresponding author \emph{E-mail address:} cphillips@anl.gov}
\affiliation{Mathematics and Computer Science Division, Argonne National Laboratory, Argonne, IL 60439, USA}
\author{Tom Peterka}
\affiliation{Mathematics and Computer Science Division, Argonne National Laboratory, Argonne, IL 60439, USA}
\author{Dmitry Karpeyev}
\affiliation{Mathematics and Computer Science Division, Argonne National Laboratory, Argonne, IL 60439, USA}
\author{Andreas Glatz}
\affiliation{Materials Science Division, Argonne National Laboratory, Argonne, IL 60439, USA}

\date{\today}

\begin{abstract}
In type-II superconductors, the dynamics of superconducting vortices determine their transport properties.  In the Ginzburg-Landau theory, vortices correspond to topological defects in the complex order parameter.  Extracting their precise positions and motion from discretized numerical simulation data is an important, but challenging task. In the past, vortices have mostly been detected by analyzing the magnitude of the complex scalar field representing the order parameter and visualized by corresponding contour plots and isosurfaces.  However, these methods, primarily used for small-scale simulations, blur the fine details of the vortices, scale poorly to large-scale simulations, and do not easily enable isolating and tracking individual vortices.  Here we present a method for exactly finding the vortex core lines from a complex order parameter field. With this method, vortices can be easily described at a resolution even finer than the mesh itself.  The precise determination of the vortex cores allows the interplay of the vortices inside a model superconductor to be visualized in higher resolution than has previously been possible.  By representing the field as the set of vortices, this method also massively reduces the data footprint of the simulations and provides the data structures for further analysis and feature tracking.
\end{abstract}


\maketitle





\section{Introduction}
Many phenomena in nature can be described by the behavior of complex scalar functions or vector fields, ranging from electromagnetic fields to director fields in liquid crystals, spins in magnets, and complex order parameters in superfluids and superconductors.  Topological defects in those functions or fields represent important features of the underlying physical system: Examples are (zero-dimensional) point defects or monopoles, (one-dimensional) defect lines or strings, and (two-dimensional) domain walls.
Here we concentrate on defect lines, which in the case of a complex scalar field are defined by one-dimensional manifolds, where the phase of the complex function is undefined. 
These topological singularities or defects are typically associated with circulations in the phase gradient and are referred to simply as vortices.  
Substantial work has been invested in studying the dynamics  of vortices in different contexts, such as crossing and reconnection and the formation of knots in superfluid vortices \cite{PhysRevLett.71.1375,SuperfluidKnots}, in light waves \cite{LeachNature}, and in fluid flows \cite{Kleckner2013}, as well as their evolution in more mathematically generalized contexts \cite{Dennis2007}.





\begin{figure*}[htbp]
   \centering
   \includegraphics[width=7in]{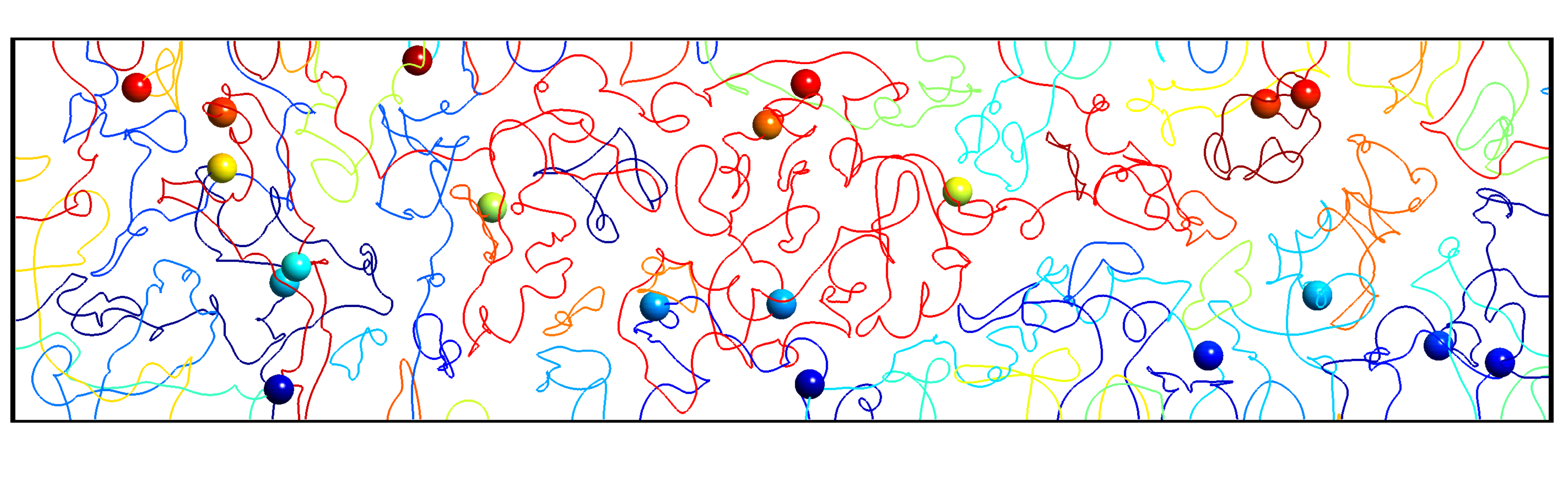}
   \caption{View along the x-axis of a superconducting material simulated by using the TDGL equations.  We show the material defects, or inclusions (spheres), and the tangled vortex loops extracted by the methods described here. The magnetic field and current along the x-axis cause the vortices to twist and writhe, and the inclusions pin the vortices in place.  The vortices were extracted from a complex scalar field discretized over a grid of 256$\times$512$\times$128 points.} \label{fig:vortex}
\end{figure*}

In type-II superconductors, an externally applied magnetic field penetrates the system above the first critical field in the form of flux tubes (vortices), which carry integer numbers of flux quanta (typically one flux quantum). 
The magnetic flux in the vortex core is screened by a circular supercurrent around it.  In the dissipative regions, vortices are dynamic objects that nucleate and annihilate; they can cut each other and reconnect.  In static situations, vortices can be pinned by material defects inside the superconductor. The behavior of vortices carrying magnetic flux determines the material's ability to sustain the dissipationless/superconducting state. When vortices move, the system becomes dissipative, and a finite voltage drop across the system is observed.  In the Ginzburg-Landau theory of superconductivity, the local superconducting properties of the material are described by a spatially dependent complex order parameter $\psi$, and vortices correspond to topological phase singularities of $\psi$ accompanied by a suppression of its magnitude.  Using the time-dependent Ginzburg-Landau (TDGL) equations, coupled partial differential equations evolving the scalar $\psi$ field in time, one can find steady-state solutions of the superconductor in the presence of external magnetic fields and applied currents. 

Simulations to model superconductors via the TDGL equations are numerically intensive.  Until recently, this method usually has been limited to 2D simulation \cite{PhysRevB.80.054506,PhysRevB.69.094521,PhysRevB.76.024509,1997Nonli..10..579C} or small 3D simulation \cite{Qiang}.  Now work has been initiated, however, to implement large 3D simulations where macroscale phenomena can be observed \cite{TDGL,PhysRevB.84.180501} taking into account the collective dynamics of many vortices.  Reaching the macroscale in these large 3D simulations requires both a stable numerical discretization of the TDGL equations \cite{TDGL} and the use of advanced computing resources.  It also requires the codesign of analysis techniques that can scale with the application. For large and long simulations, recording the state of the system by frequently storing the entire state of the system will be untenable.  Fortunately, in order to support a detailed analysis of the vortex dynamics over time, only the locations of vortices themselves are required.

Here we introduce a data analysis method for the numerical extraction of a vortex from a complex order parameter field obtained from large-scale simulations of a type-II superconductor.  This analysis generates vortex objects, or reduced mathematical representations of one-dimensional curves that correspond to individual vortices from a discretized complex scalar field.  An example of the complex and tangled vortex state that can extracted with this method is shown in Figure \ref{fig:vortex}. This analysis has applications to discretized complex fields containing topological defects, for example, optical vortices in electromagnetic fields as well as other problems described by the complex Ginzburg-Landau equations such as screw dislocations \cite{PhysRevLett.80.1770} cosmic strings \cite{0034-4885-58-5-001}, superfluidity, and Bose-Einstein condensation; strings in field theory \cite{RevModPhys.74.99}; topological defects in liquid crystals \cite{spiral}; and models of fluid dynamics with complicated nonlinear dynamics \cite{PhysRevLett.96.074501}.

In Section \ref{sec:prior}, we briefly survey prior methods for detecting vortices in complex scalar fields.  In Section \ref{sec:method}, we provide our algorithm.  
We show how vortex core points are detected, interpolated, and efficiently stitched together to form topologically ordered objects and then further compacted into mesh-independent objects. In Section \ref{sec:performance}, we discuss the performance and scaling of this algorithm with respect to the mesh size or the density of the vortex state.  In Section \ref{sec:conclusion}, we provide concluding remarks.




\begin{figure*}
  \begin{minipage}[c]{0.67\textwidth}
    \includegraphics[width=\textwidth]{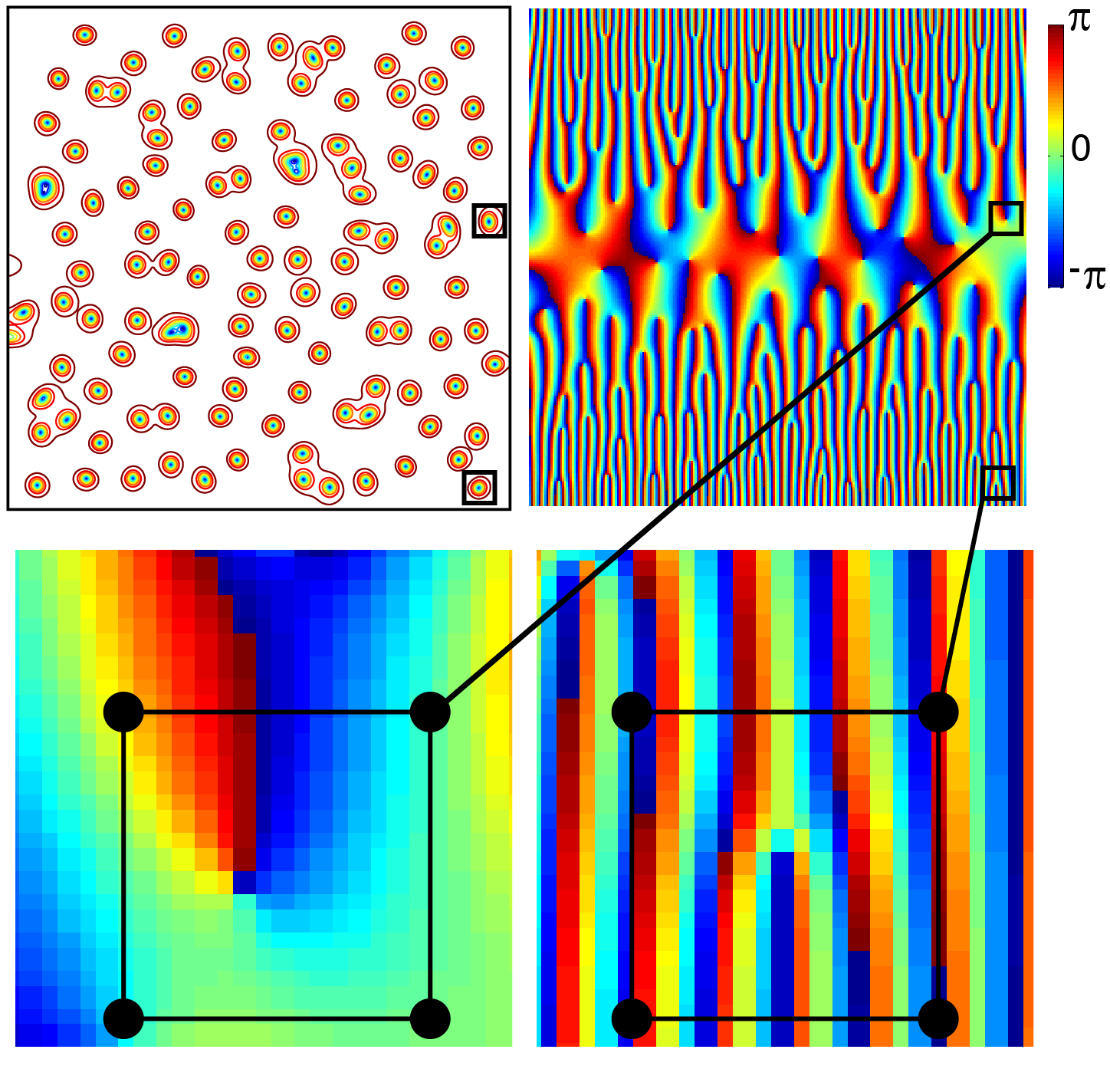}
  \end{minipage}\hfill
  \begin{minipage}[c]{0.3\textwidth}
    \caption{
(top left) The contour plot of a slice of the magnitude of a complex field.  (top right) The plot of the phase of $\psi$ for a slice of the complex field.  A black box is drawn around two vortex cores in both slices.  (bottom left) For the vortex core in the middle of slice, integrating the phase around the box shows a phase jump.  (bottom right)  For the vortex core at the bottom of the slice, integrating the phase around a box of the same size will produce errors because the phase oscillates four times along the top and bottom edge of the box.  The region of the slice where the phase lines become crowded is arbitrarily determined.  By applying a gauge transformation at each point, locally the data can be transformed to have the lowest possible density of phase lines anywhere in the slice.
    } \label{fig:contourphase}
  \end{minipage}
\end{figure*}


\section{Background and Prior Work} \label{sec:prior}
In terms of $\psi$, a vortex line is defined as the locus of points where $|\psi|$ = 0 and where $\oint \! \nabla \theta \cdot ds =  2n\pi$, where $\theta$ is the phase of $\psi$.  The integration is performed on a closed loop around a vortex line, and $n$ is a nonzero integer, usually $\pm 1$.  The sign of $n$ indicates the chirality of the vortex with respect to the direction of integration around the closed loop. Figure \ref{fig:contourphase}, which shows the magnitude and phase of $\psi$ in a $yz$ plane slice of a 3D field, demonstrates the correspondence between these two measures. Two black boxes surround two vortex cores on both the top left and right images.  In the top left image, the contour lines indicate that $|\psi|$ = 0 in the center of the boxes.  For the right image, the expanded views at the bottom show the defect in the phase field present at both locations.  In both cases the phase sums to 2$\pi$ in an appropriately defined loop. 

In numerical studies of type-II superconductors, the phase information of the field is typically disregarded, and vortices are identified by examining the contour plots of $|\psi|$ in 2D \cite{1997Nonli..10..579C,PhysRevB.76.024509} (or the isosurfaces of $|\psi|$ in 3D \cite{Qiang}).  Sometimes the contour plot is supplemented by examining plots of the phase of $\psi$ \cite{PhysRevB.80.054506} when unusual features, such as a giant vortex state, are suspected.  The assessment of the vortex positions in these contour fields is qualitative but sufficient to show how vortices self-organize in small simulations.     

In large-scale 3D simulations, generating isosurfaces is not a viable technique for understanding vortex behavior.  First, qualitative assessments of how the vortices self-organize fails for large 3D data sets with densely packed entangled vortices.  Second, storing data to visualize a contour or isosurface does not significantly reduce the size of the data in a time step.  Third, the format of isosurfaces and contour data, especially in dense distributions of vortices, does not easily lend itself to tracking individual vortex dynamics over time; more precise numerical interpretations are required.  Fourth, using contours to find vortices fails completely when the superconductor model includes simulated material defects (shown in Figure 1) often modeled as a suppression of the magnitude of the  $\psi$ field \cite{TDGL}. With an isosurface method, the location of a vortex core inside an inclusion cannot be visualized because the magnitude of the field around the vortex is suppressed inside the inclusion.

Here we introduce a data analysis method for exact numerical extraction of a vortex from a complex order parameter field obtained from large-scale simulations of a type-II superconductor.  Rather than relying on the contours of the magnitude of the complex field, our analysis method finds the curves of singularity points in the phase of $\psi$ by integrating the phase of $\psi$ around small loops.  The analysis then extracts these points in topologically ordered sets that represent each vortex.  This method also allows direct measurement of the chirality of a vortex, or the direction (clockwise or counterclockwise) of the supercurrent flow around the vortex core line.   This method reduces the representation of a 3D field to a set of discrete 1D objects.  Previously, parts of these techniques have been applied to trace vortices in small-scale 3D type-II superconductor data \cite{refId0,PhysRevB.67.144514} and to find optical vortices in experimentally measured electromagnetic fields \cite{1464-4258-11-9-094020,1464-4258-6-5-009}. However, the target and scale of our application, the techniques for unwrapping the phase locally, the interpolation to more precisely describe the vortex object, the method for rapidly constructing a vortex object from a subgraph, the introduction of a compact and mesh-independent representation, and the general consideration of the computational efficiency of the extraction are unique to our work.

\section{Method} \label{sec:method}
 The source of our data set is a TDGL model implemented on a structured finite-difference discretization mesh with a uniform grid spacing oriented along the $x$,$y$,$z$ axes of the space.  We refer to this as a regular Cartesian mesh.

\begin{algorithm}
\caption{Vortex Feature Detection}
\label{alg:overall}
\begin{algorithmic}[1]
\State{Test each mesh element face to see if it is punctured by vortex. (\ref{sec:puncture})}
\State{If desired, for all punctured faces, interpolate the location of the puncture point.  Otherwise treat as the center of the face. (\ref{sec:interp})}
\State{For each punctured face, add nodes and edges to subgraph. (\ref{sec:graph})}
\State{Trace each vortex through the constructed subgraph to segment and order the set of vortex points into separate vortex structures. (\ref{sec:partition}) }
\State{Fit curves through the ordered sets of vortex points. (\ref{sec:compact})}
\end{algorithmic} \label{algorithm}
\end{algorithm}

Our algorithm, as described in Algorithm \ref{alg:overall}, extracts vortices from the data by performing closed loop integrations of the phase around every mesh element face.   The integration is discretized over the four edges of the mesh face, using the values at the four corners.  In Figure \ref{fig:contourphase}, one can immediately see an issue with this scheme.  While even a large loop around the vortex core on the bottom left unambiguously encircles a defect in the phase field and the phase increments will sum to 2$\pi$, only a very small loop, perhaps even smaller than the resolution of the mesh, can be used on the bottom right.  Otherwise, the phase changes by more than $\pi$ along individual segments, and using only the value at segment endpoints will result in error.  In Section \ref{sec:puncture}, we show how this problem is corrected by applying a gauge transformation along the path of integration.  

If a vortex passes through a mesh element face, we say it ``punctures'' the face, and the exact point it penetrates the face is the ``puncture point.''    When a mesh face is found to be punctured, an interpolation can be applied, based on the values of $\psi$ at the grid points of the mesh, to determine where inside the face $|\psi|$ = 0, or the unique location where the vortex punctures the face.   In Section \ref{sec:interp}, we provide a generalized technique for finding the puncture point inside a generalized mesh element face.

In order to facilitate the topological reconstruction of each vortex, the information determined in Step 1 is used to construct a graph, described in Section \ref{sec:graph}.   In Section \ref{sec:partition}, we show how this graph, which is a subgraph of the mesh,  can be rapidly traversed to reconstruct each vortex core line, as well as used to identify rare points of contact between vortices.  In Section \ref{sec:compact}, we show how the representation of the vortex core line can be made compact and mesh independent.

\subsection{Finding Punctured Faces} \label{sec:puncture}
\begin{figure}[htb]
        \centering
    \includegraphics[width=3in]{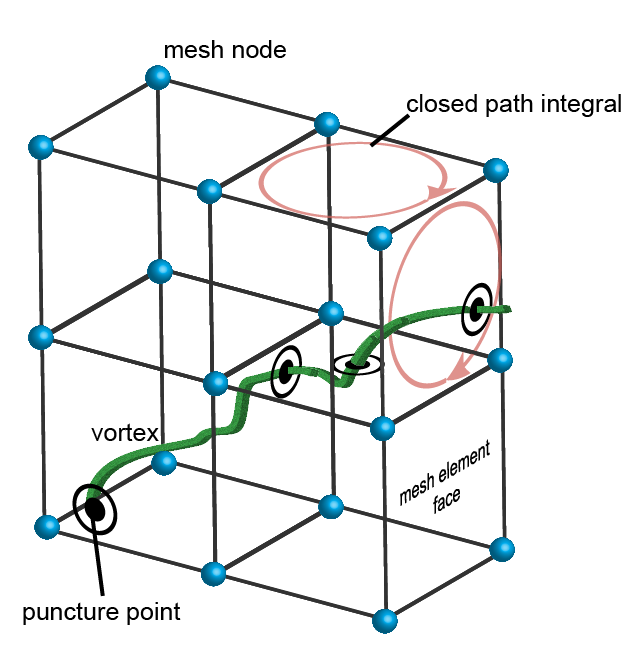}
    \caption{Illustration of a vortex line weaving through four mesh elements.  Blue balls represent grid points where the value of $\psi$ is known.  The bullseyes indicate the four puncture points.  Of the two integrals along closed paths illustrated, one has a value of zero and one has as a value of one.}
    \label{fig:illustration}
 \end{figure}
 
Given a set of complex values $\psi$ that have been calculated on each point of a mesh, vortex lines can be localized by calculating the integral 
\begin{equation}
n=-\tfrac{1}{2\pi}\oint \! \nabla \theta \cdot dl 
\end{equation}\label{eq:int}
around closed paths in the mesh.   When the value of $n$ is a nonzero integer (usually $\pm1$), then the path encircles a vortex line, and the sign of $n$ indicates the chirality of the vortex with respect to the face normal.  The smallest closed path that can be calculated is a noncolinear triangle of points, such as half a mesh element face.  For simplicity, however, we perform closed paths integrals around the perimeters of the rectangular mesh faces.   The closed path  integral is broken up into a sum of line integrals calculated over each line segment of the path.   An illustration for mesh elements is provided in Figure \ref{fig:illustration}, or
 \begin{equation}
n\equiv -\tfrac{1}{2\pi}\sum_1^m \Delta{\theta}_{i,i-1} \label{eq:sum},
\end{equation} 
where 
\begin{equation}
\Delta\theta_{i,i-1} = \textrm{mod}(\theta_i - \theta_{i-1}  +\pi,2\pi)-\pi
\end{equation}
 and $m$ is the number of segments defining the path around the face.
 
 The value of the phase of $\psi$ at each grid point is stored in an $n_z\times n_y \times n_x$ 3D array $\Theta$, where $n_i$ is the number of grid points along the $i$th axis. 
 In order to calculate the phase differences in the $x$, $y$, or $z$ direction, a copy of $\Theta$ is rolled in the axial direction, that is, circularly shifted one index position, subtracted from $\Theta$, and the 2$\pi$ modulo is taken of the resultant multidimensional array.  We use the notation $\Theta_{1,0,0}$, $\Theta_{0,1,0}$, and $\Theta_{0,0,1}$ to represent the $\Theta$ matrix rolled in the positive $x$, $y$, and $z$ direction, respectively.

\begin{figure}[htb]
        \centering
    \includegraphics[width=2in]{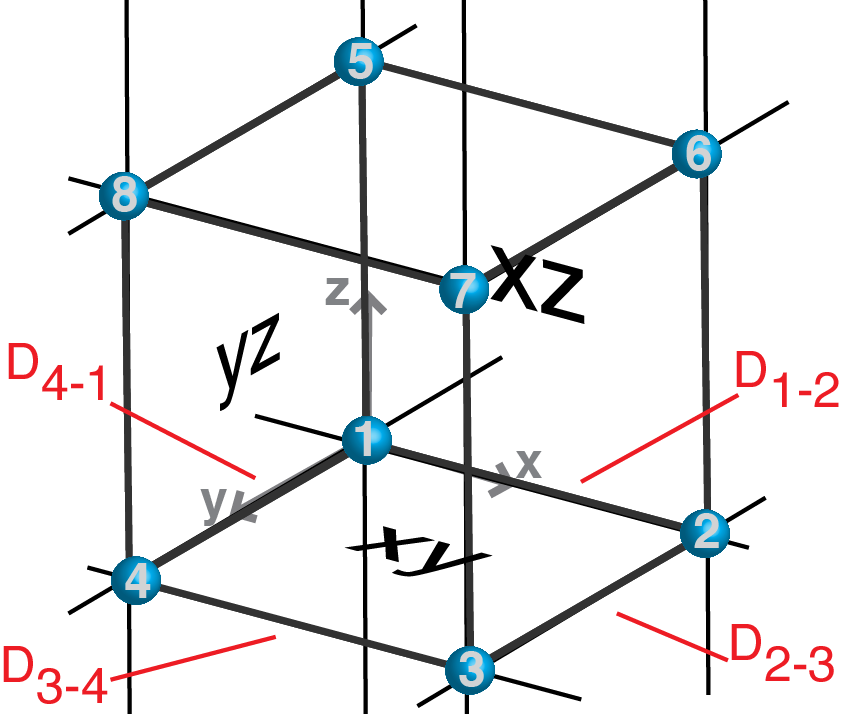}
    \caption{One mesh element in the grid.}
    \label{fig:oneelement}
 \end{figure}
 
For example, Figure \ref{fig:oneelement} shows an annotated illustration of a single mesh element. We let $D_{1-2}$ equal the 2$\pi$ modulo of $\Theta_{1,0,0} - \Theta$.     Likewise, $D_{4-1}$  = $\Theta- \Theta_{0,1,0}$.  Therefore $D_{3-4}$ and $D_{2-3}$ are constructed by applying a circular shift to $D_{1-2}$ and $D_{4-1}$, respectively, in the $y$ and $x$ axis, respectively.  The sum of these four arrays, $n_{xy}$, is a 3D array containing the integration of the phase, or a calculation of Equation (2), around the perimeter of every mesh element face in the $xy$ plane.   

In the remainder of this section we explain two corrections that make the calculation valid over the entire simulation space.

%
%

This integration calculation, broken over the four segments of the mesh element face, is an acceptable calculation of the contour integral as long as the phase of $\psi$ does not change by more than $\pm\pi$ along any line segment.  
In a TDGL simulation, however, the gradient of the phase of $\psi$ depends on the vector potential $\bf{A}$ and the applied current. 
In Figure \ref{fig:contourphase}, for example,  the box drawn around the vortex core at the bottom of the plot has many wrappings of the phase along the top and bottom edges of the box, meaning the phase changed by $\pi$ several times along the segment.  
If the contour integral was performed around an arbitrarily small path around the vortex, or if the value of $\psi$ could be sampled at arbitrarily small line segment intervals along the contour, the calculation would be correct.  
However, the resolution of our calculation is determined by the resolution of the structured mesh.   
Nonetheless, the value of $\psi$ can be locally transformed to make the calculation valid again.  
The phase of the order parameter in the TDGL model is not uniquely defined; it depends on the choice of the gauge for the vector potential.
This choice of gauge determines where in the plot of the phase of $\psi$ of Figure \ref{fig:contourphase} the phase lines are dense (the top and bottom) and where they are not (the middle).   
By applying gauge transformations along the contour integral, which changes the vector potential such that high-frequency oscillations of the phase of $\psi$ are removed locally, a unique vortex detection and highest precision interpolation are possible.     
In Appendix A, we derive a gauge-invariant contour integral. 
The result of this calculation is a set of multidimensional arrays that are added to the phase difference multidimensional arrays. 

In order to perform the integration loop correctly at the boundaries of the simulation data, the correct boundary conditions need to be applied. Three types of boundary conditions are possible in a TDGL simulation. The first is the open or ``no current"  boundary condition; in this case, nothing needs to be done.  
The second and third types are ``periodic" and ``quasiperiodic,'' respectively.  
In both these cases, the end faces of the mesh are connected to each other.  The mesh includes an extra slab of mesh element that straddles the two end faces. If the boundary condition is periodic, the calculation performed on this extra slab is no different from anywhere else.   
Depending on the choice of vector potential, the periodic boundary conditions in one direction must be replaced by a ``quasiperiodic" boundary condition.
In this case, the magnitude of the order parameter is still periodic, but its phase shifts across the boundary. 
The integration around a mesh element face straddling a quasiperiodic boundary requires a correction term for this phase shift where the boundary is crossed.   In Appendix B, the calculation for the quasiperiodic boundary condition correction is provided. 
The result of this calculation is a two-dimensional array that is added to a two-dimensional slice of the phase difference array when applicable.

We have shown how all the contour integrals around all the mesh faces can be described by a series of circular shifts, additions and subtractions for the regular data pattern of a  structured mesh.  In practice, in order to keep the memory footprint of the problem small, the operations can be performed on slices of the 3D array.  The regular and local nature of these calculations can be optimized in various ways to maximize data reuse, memory, and parallelism of a given computational algorithm.  

\subsection{Interpolating within a Mesh Element Face} \label{sec:interp}

\begin{figure*}[htb]
        \centering
    \includegraphics[width=7in]{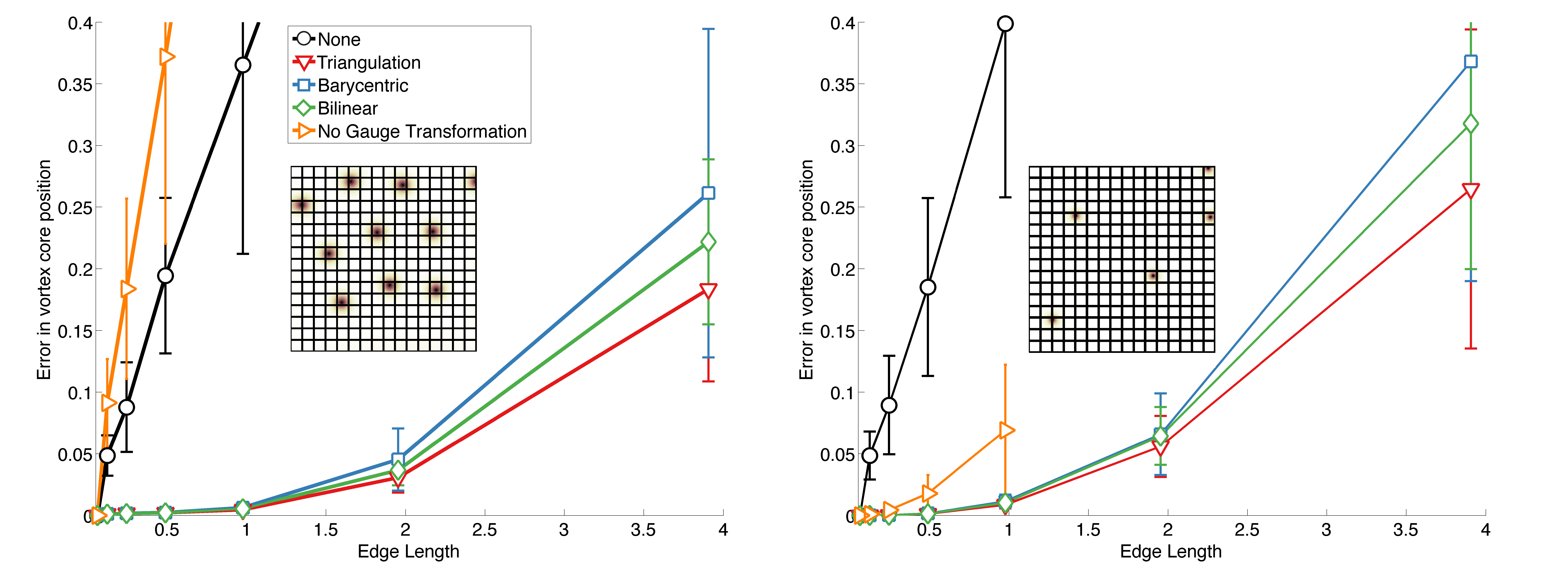}
    \caption{The precision of different interpolation methods for a dense (left) and sparse (right) vortex core distributions in a 2D plane.  For comparison, the result of the interpolation if the gauge transformation is not applied to the data (only shown over the range where the correct number of vortices was detected) is also included.  All units are coherence length, or the length unit of the simulation.  Inset in each plot is an example 1/16th of the 2D plane for each case, showing ten and five vortex cores, respectively.  }
    \label{fig:Interp}
 \end{figure*}
 
Given a punctured face, a more precise prediction of the puncture point can be determined by interpolating from the values of $\psi$ on the four grid points of the face.   Here we use the other definition of a vortex core point, a point where $|\psi| = 0$, or both the real and imaginary component of $\psi$ are zero.  Given the four $\psi$ values, we predict where in the interior of the face $\psi$ = 0.  This is significantly more computationally expensive than calculating the contour integral around a face, and thus it is not generally used as the test to predict whether a face is punctured.   

In Appendix C, three methods are provided for interpolating the puncture point: triangulation, inverse bilinear interpolation, and inverse barycentric interpolation.   In Figure \ref{fig:Interp}, the precision error inherent in these three methods is shown for both a dense and a sparse configuration of 2D vortices.  The mean error in predicting the position of the vortex core point is compared with the length of the side of a mesh element (both in units of $\xi_0$, the zero temperature coherence length, the physical length unit used in simulation).  The three methods are compared with assuming that the vortex core center is at the center of the punctured face (None).  The grids in the top of Figure \ref{fig:Interp}  correspond to the coarsest edge length of 3.9.  For this data, triangulation is slightly superior to inverse bilinear interpolation and inverse barycenteric interpolation, but all perform similarly.  At the standard edge length chosen in simulation, 0.5, all three have an error that is less than 1\% of the edge length.  

Applying the gauge transformation not only makes the contour integral numerically valid in dense vortex systems but also significantly improves the prediction of the position of the vortex core point.  The impact of not applying the gauge transformation (and interpolating with the triangulation method) is shown in both plots.  Data is shown only over the range where the correct number of vortices was identified.  In the dense configuration, this method performs worse than using the gauge transformation with no interpolation, because vortices are sometimes not found in the correct grid cell.  In the sparse configuration, we see that although the gauge transformation is not necessary to find punctured element faces for sufficiently small mesh elements, not applying the gauge transformation to the data adds significant error to the interpolation.

\subsection{Constructing a Graph Structure} \label{sec:graph}
The 4D array $n$ for each planar contour integral contains only 0, 1, or -1, where the nonzero elements of $n$ corresponds to the punctured mesh element faces.  The sign of the nonzero element corresponds to the chirality of the vortex relative to normal axis of the face it is puncturing.  

In reference \cite{opticaldiss}, the set of puncture points associated with the nonzero faces in $n_{xy}$, $n_{xz}$, and $n_{yz}$ were compacted into a list and then topologically sorted by Euclidean distance to partition them into separate vortex objects.  Optimally implemented, this algorithm has a computational complexity of $\bigO(N\textrm{log}(N))$, where $N$ is the number of points.   However, using a Euclidean distance criterion to sort points can produce incorrect results.   In theory, two vortices that do not puncture the same mesh elements can have points arbitrarily close together.  Also, this method does not extend well to meshes where mesh elements are not uniformly sized cubes.  For heterogeneous and irregular meshes, no simple distance criterion will work.  Instead, we propose a scheme that retains the connectivity information of the puncture points that is implicit in the mesh structure, allows fast reconstruction of the vortex objects, and is of computational complexity $\bigO(N)$.  

\begin{figure}[htb]
        \centering
    \includegraphics[width=3.5in]{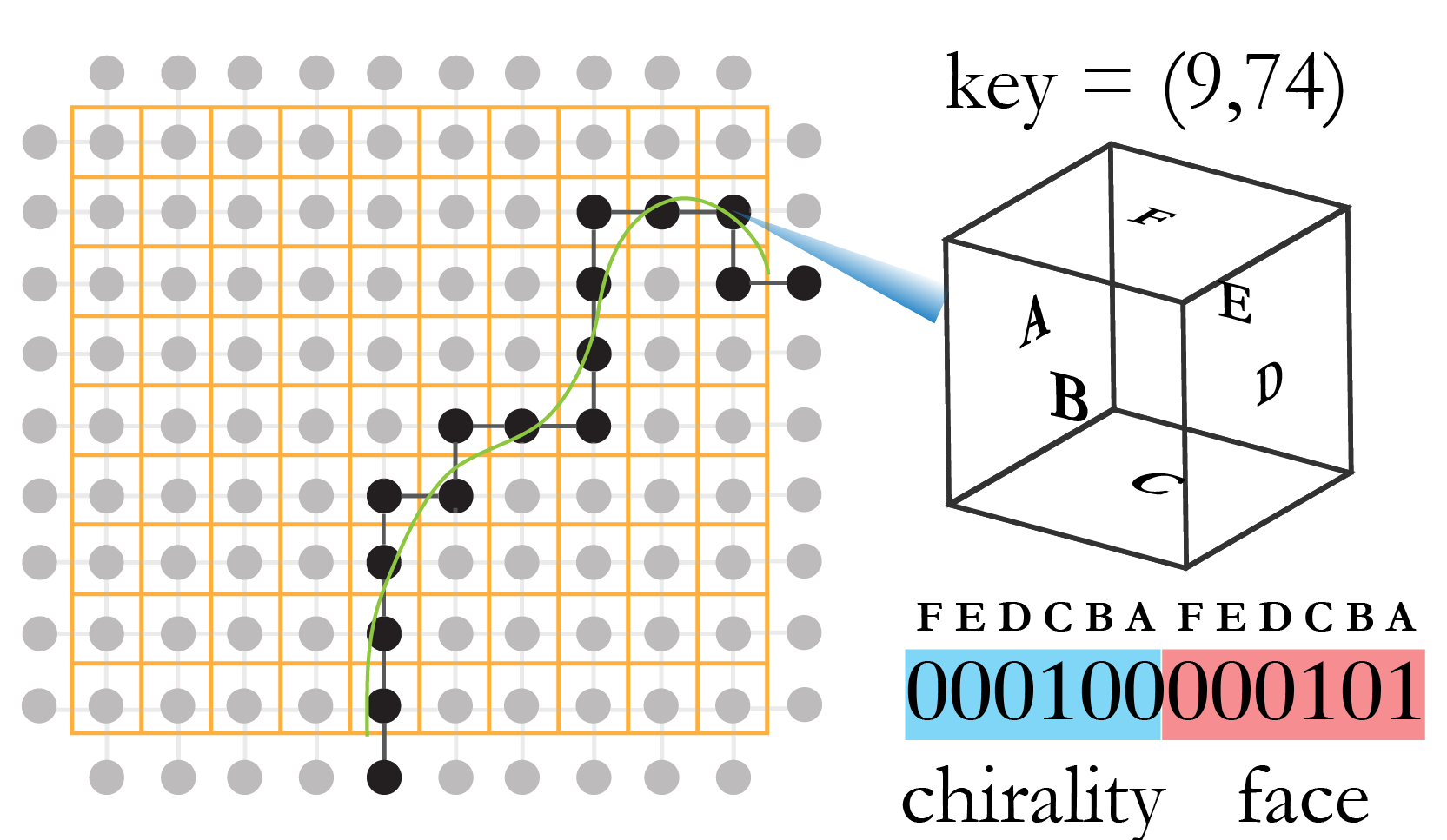}
    \caption{Left: Illustrated in 2D, a mesh can be interpreted as a graph structure.  The path of a vortex (green) puncturing the mesh can be represented as a subgraph of this graph.  Right: For each punctured mesh element, the subgraph dictionary stores a 12-bit number (i.e., a node) that indicates which faces were punctured (i.e., the edges) and the chirality of the vortex puncturing the face.  In this example, faces A and C were punctured; the vortex has a negative chirality relative to face A and a positive chirality relative to face C.}
    \label{fig:Cdiagram}
 \end{figure}

One way to interpret the structure of a mesh is as a graph, where mesh elements are nodes and mesh faces are edges connecting two mesh elements.  We assume that given a mesh element, there is a fixed way to order its faces and that the identity of each mesh element neighboring the original element via each face is accessible via an $\bigO(1)$ calculation, either because of the regular structure of the mesh or through a precalculated look-up table.  The mesh elements and mesh faces punctured by a set of vortices are then a subgraph of this graph. The nodes of the subgraph are punctured mesh elements.  The edges of the subgraph are the shared punctured faces of neighboring mesh elements. This is illustrated in 2D on the left in Figure \ref{fig:Cdiagram}.  Both constructing the subgraph and connecting the core points by tracing paths through the graph are $\bigO(N)$ calculations, where $N$ is the number of core points, that is, punctured faces.

The subgraph structure can be constructed simultaneously with finding core points by adding an edge each time a punctured face is found.  Since the fraction of mesh elements that are punctured is very small even in a dense configuration of vortices, we choose to use a dictionary, or hash table, to store the nodes and edges.  On average, inserting, retrieving, and deleting a key-value from a hash table are $\bigO(1)$.  The key is an integer that uniquely identifies a mesh element, or the node.  The value is a binary string representing the punctured faces of the element, or the edges of the node.  The chirality of each vortex face puncturing is also stored in a second binary string.  Thus for each nonzero element of $n$, two nodes, the two mesh elements that share the punctured face, are added to the dictionary (if not already present), and an edge is added connecting the nodes.  The interpolated vortex center coordinates are stored in a separate dictionary by using a key that uniquely represents the face.  

 For a mesh with hexahedral elements, each node can have edges to only six other nodes, so the edges can be represented as a 6-bit string.  In a regular Cartesian mesh, no look-up table between elements and connecting faces is required, because of the simple structure of the mesh.  As shown in Figure \ref{fig:Cdiagram} on the right, the key for each punctured mesh element is its unique coordinate position in integer index space.  The value stored for each key is a 12-bit string, where bits 0-5 are set if faces A-F are punctured, and bits 6-11 indicate the chirality of the vortex puncture.   For the chirality bits, a bit has meaning only if the associated face bit is set.  A value of 0 indicates the more common positive chirality, while a value of 1 indicates a negative chirality \footnote{It is possible to devise a slightly more compact scheme since there are only $3^6$ independent states.}.
 
This algorithm can be trivially extended to an unstructured mesh, albeit dependent on the availability of a look-up table for determining what face connects which elements.  Neighbor element lookup is commonly supported in meshing libraries, such as libmesh \cite{libMeshPaper}.  

\subsection{Tracing Each Vortex to Extract the Topological Structure} \label{sec:partition}
In the subgraph, each vortex maps to a set of connected nodes. In order to extract the topologically ordered set of puncture points that define each vortex, a node is acquired from the subgraph dictionary, and its edge information is used to acquire the next node in a chosen direction.  Each node is removed from the dictionary upon acquisition.  This procedure is repeated until no more nodes are found.  The procedure is repeated for the other direction of the original node, and the two lists of nodes are appropriately concatenated. This ordered list of nodes represents a complete vortex and can be converted back into an ordered list of puncture points.   If the interpolated puncture points were stored in a dictionary, then their key can be reconstructed from the nodes in the list, and each face point can be replaced by the higher-precision interpolated point. 
To find all the vortex objects, we acquire and trace the nodes until the dictionary is empty.  Using this subgraph dictionary, we construct the set of vortex objects in computational time linear to the number of puncture points in the system.


If two vortices puncture the same mesh face, this cannot be resolved.  The algorithm here depends on the assumption that mesh data is generated at a resolution that is commensurate with the interaction lengths of meaningful physical processes.  However, in extremely rare cases --- far less than 0.1\% of the punctured mesh elements --- two vortices can be close enough to puncture the same mesh element but not the same face.  Even though technically the two vortices may not be connected, for the purpose of analysis they are treated as a single vortex object. If during the trace a node with connectivity $>$ 2 is found, that is, with more than two face bits set, then new traces are initiated in each face bit direction (barring the direction of the original trace), and the algorithm returns a set of lists of ordered points, one for each trace direction and one containing just the points of the high-connectivity node.       

In an even rarer case, a vortex could be close enough to an edge or corner of a mesh element such that a contour integral interprets the vortex as penetrating zero, two, or three faces. The likelihood of this happening is directly related to the precision of the calculation of the contour integral.  In an infinite precision calculation, this event has zero probability of occurring.  In a single- or double-precision calculation, the probability is still extremely low.  In fact, we have not observed this statistically unlikely event yet.  Rather than adding additional expensive checks to the vortex core finding or tracing, this case would best be detected by checking traced vortices for anomalous properties (e.g., having an end that does not terminate in a boundary).   Note that this cannot occur because of precision error in the interpolation, since even if a vortex core is interpolated to be slightly outside of a face, it is still treated as puncturing the original face.  

\subsection{Creating a Compact Mesh-Independent Vortex Object}\label{sec:compact}
At this stage of the algorithm, a vortex object is represented by an ordered set of puncture points.  The number of puncture points is determined by the mesh resolution.  Commonly, vortices are nearly straight curves that span one dimension of the mesh; thus, a far more compact, and even mesh-independent representations of each vortex is possible.  Here we discuss one method for compacting the vortex representation.

If we ignore the wrapping of a vortex across periodic boundaries, by, for example, cutting a vortex into pieces when it wraps or creating an unwrapped vortex using periodic images of the vortex, then a vortex represented as an ordered list of puncture points is a polyline.  Polyline simplification, or the decimation and curve fitting of a polyline to create a more compact representation, is a well-studied problem in computer graphics with numerous available algorithms.  Here we decimate our polyline using the Ramer-Douglas-Peucker (RDP) algorithm \cite{rdp,Ramer1972244}, and then further reduce and fit the polyline using Schneider's algorithm \cite{fitcurves}.

Given a polyline, RDP reduces it to a simpler polyline by recursively dividing it until a distance criterion is met by each segment.  Schneider's algorithm fits piecewise cubic Bezier curves to a polyline again, by dividing the polyline until a distance criterion is met by each curve.  Each piecewise cubic Bezier curve is represented by two endpoints and two control points.  It is not strictly necessary to apply RDP to a polyline before applying Schneider's algorithm; however, the cost of decimating the polyline and evaluating the distance criterion is cheaper for RDP than for Schneider's algorithm, and thus, this prestep modestly improves the net time of polyline simplification. While the performance of both algorithms is in worst case $\bigO(n^2)$, where $n$ is the number points, on average it is $\bigO(n\textrm{log}(n))$.  

Both RDP and Schneider's algorithm require an error parameter in units of distance for evaluating their distance criterion.  The smaller the error parameter, the truer the final piecewise curve will be to the original set of points, the larger the number of piecewise curves that represent the vortex object, and the more recursive iteration steps will be required to fit the curves.  In units of coherence length, we chose $\epsilon = 0.05$ and $0.01$ for RDP and Schneider's algorithm, respectively.  These parameters decimate the original polyline vortex by approximately a factor of 10 and then 3, when performed in series.  The final representation of the vortex object is mesh independent because, presuming the original mesh was detailed enough to capture the features of the vortex, then using finer meshes should not significantly change the final compact representation of the vortex.


\begin{figure}[htb]
        \centering
    \includegraphics[width=3.5in]{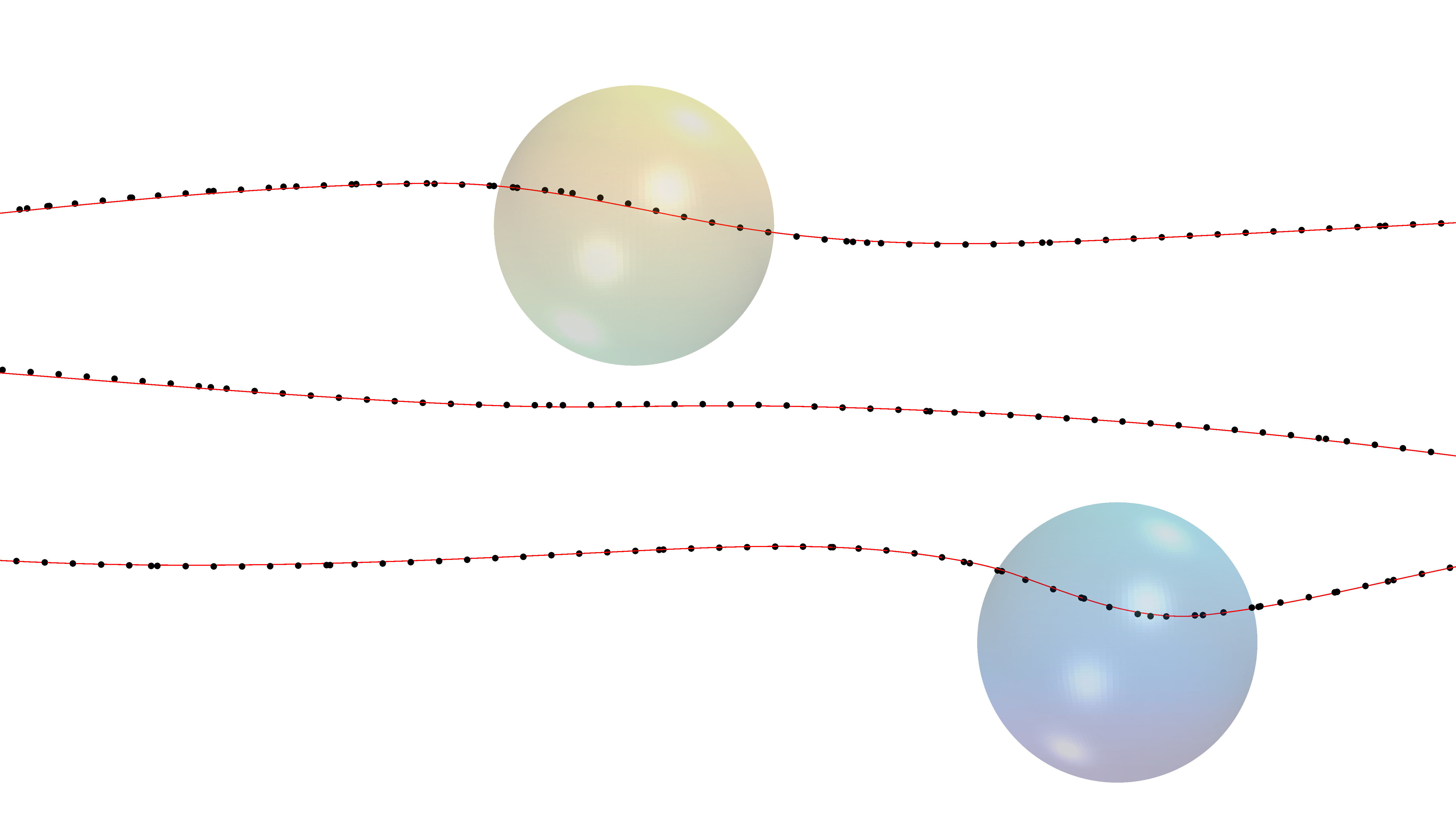}
    \caption{Three vortices, two pinned on inclusions, are shown. Black dots are puncture points. Red curves represent the piecewise cubic Bezier curves fit through the puncture points.  The details of how each vortex flexes as it traverses an inclusion are apparent.   }
    \label{fig:large}
 \end{figure}

\section{Peformance} \label{sec:performance}

A prototype version of the vortex-finding algorithm described above was implemented in Python using the \emph{numpy} library and serially on a single thread.
All benchmarks shown were performed on an Intel Core i7, 2.3 GHz with 4 cores and 16 GB of RAM.  

For a benchmark testing of the analysis code, we created a 512 MB 256x512x512 data set with a dense distribution of vortices that is periodic in the x-direction.  The data set contains 305 vortices.  However, each vortex wraps through the periodic x-boundary four times on average.  If we count each time a vortex wraps through the box individually, the data set contains 1,297 vortices (Figure \ref{fig:large}).  The total amount of time to extract all the vortices is between two and three minutes, depending on the interpolation method used. 

Table \ref{table} lists the timings of the major steps of the algorithm.  Because of the dense vortex state of this data set, performing the interpolation  and fitting the cubic Bezier curves require the largest fraction of time, nearly three-quarters of the calculation time.  Strictly speaking, both interpolating and curve fitting are optional.  Without them, a less smooth vortex object composed of ordered points is still constructed by the analysis.  We provide the timings for four different versions of interpolation (each discussed in more detail in Appendix C).  The timing difference among the methods varies by less than a factor of 2.  The triangulation method is the most computationally efficient. If we assume, however, that we are performing triangulation in a rectangle arbitrarily oriented in space, a more general case, then the efficiency drops significantly.  The inverse barycentric interpolation, which makes no orientation assumptions, is nearly as efficient as triangulation. The inverse bilinear interpolation is the most computationally expensive.  Generating and tracing the graph to construct topologically ordered vortex structures require only 8\% of the total calculation time.  Unaccounted-for time is primarily I/O operations.  
\begin{figure}[htb]
        \centering
    \includegraphics[width=3.5in]{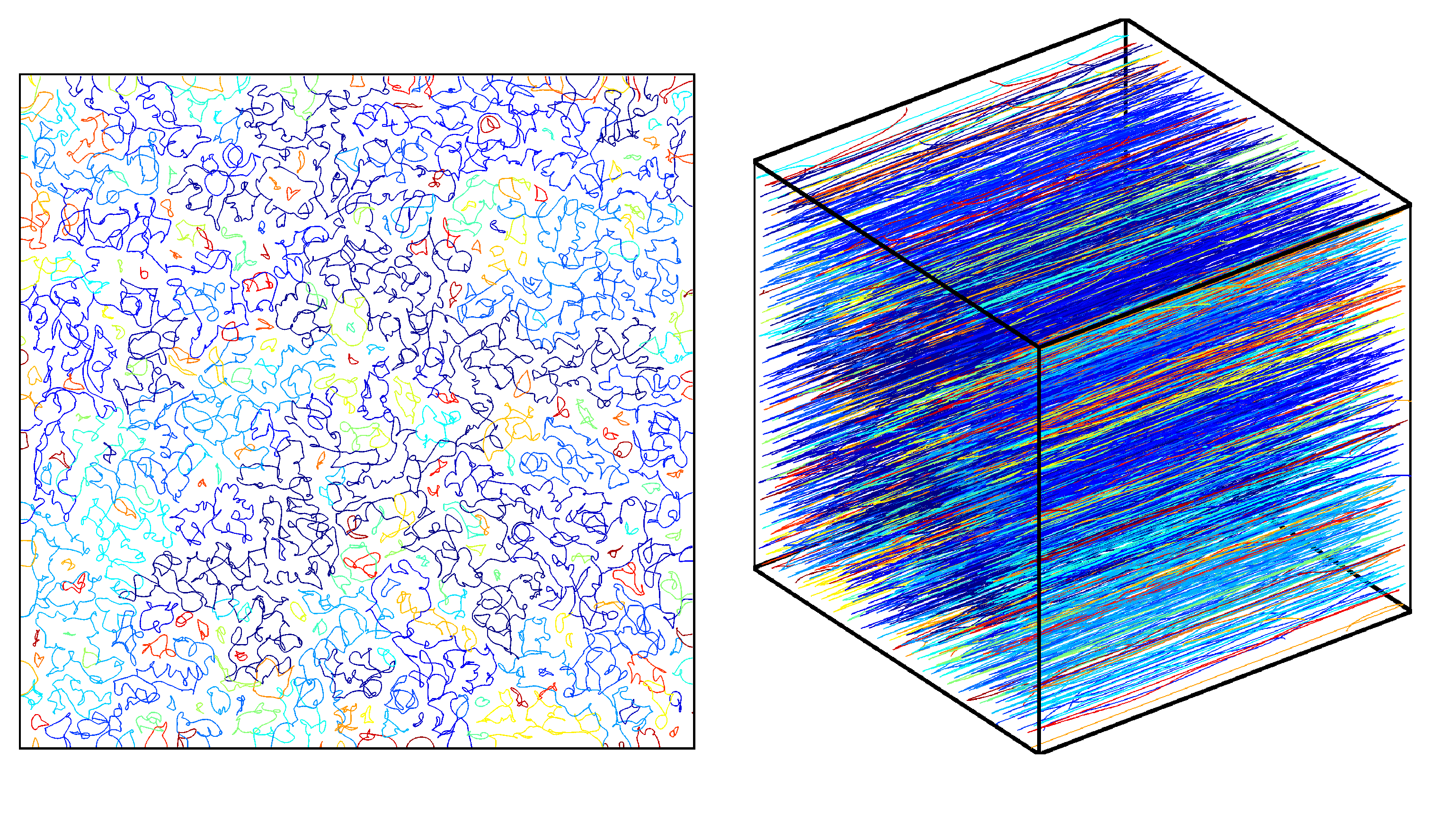}
    \caption{Benchmark data set of 256x512x512 grid points and 1,297 vortices}
    \label{fig:large}
 \end{figure}
 
 \begin{table}[htb]
\centering
\caption{Timing of algorithm for 256x512x512 grid points and 1,297 vortices}
  \begin{tabular}{  | l | c | }
    \hline
    Algorithm Step & Time (sec)  \\ \hline
    Find Punctured Faces & 23.2  \\
    Interpolation  - Triangulation & \hspace{7mm} 31.2  \\
        Interpolation  - Barycentric & \hspace{7mm} 36.3  \\
    Interpolation  - Generalized Triangulation & \hspace{7mm} 51.0  \\
    Interpolation  - Bilinear & \hspace{7mm} 52.1  \\
    Generate Subgraph and Trace Vortices & 10.8  \\
    Fit Cubic Bezier Curves & 62.0 \\
    Total (with Triangulation) & 131.2   \\

    \hline \hline
  \end{tabular}\label{table}
\end{table}

This algorithm has two important scaling dimensions:  scaling with increasing data (larger grid size) and scaling with increasing vortices.  To separate how the algorithm scales independently with respect to these two dimensions, we consider two tests.  In the first, Section \ref{scalegrid}, we keep the number of vortices fixed while increasing the grid size.  In the second, Section \ref{scalevortex}, we keep the grid size fixed while increasing the number of vortices present.

\subsection{Scaling with Grid Size} \label{scalegrid}

\begin{figure}[htb]
        \centering
    \includegraphics[width=3.5in]{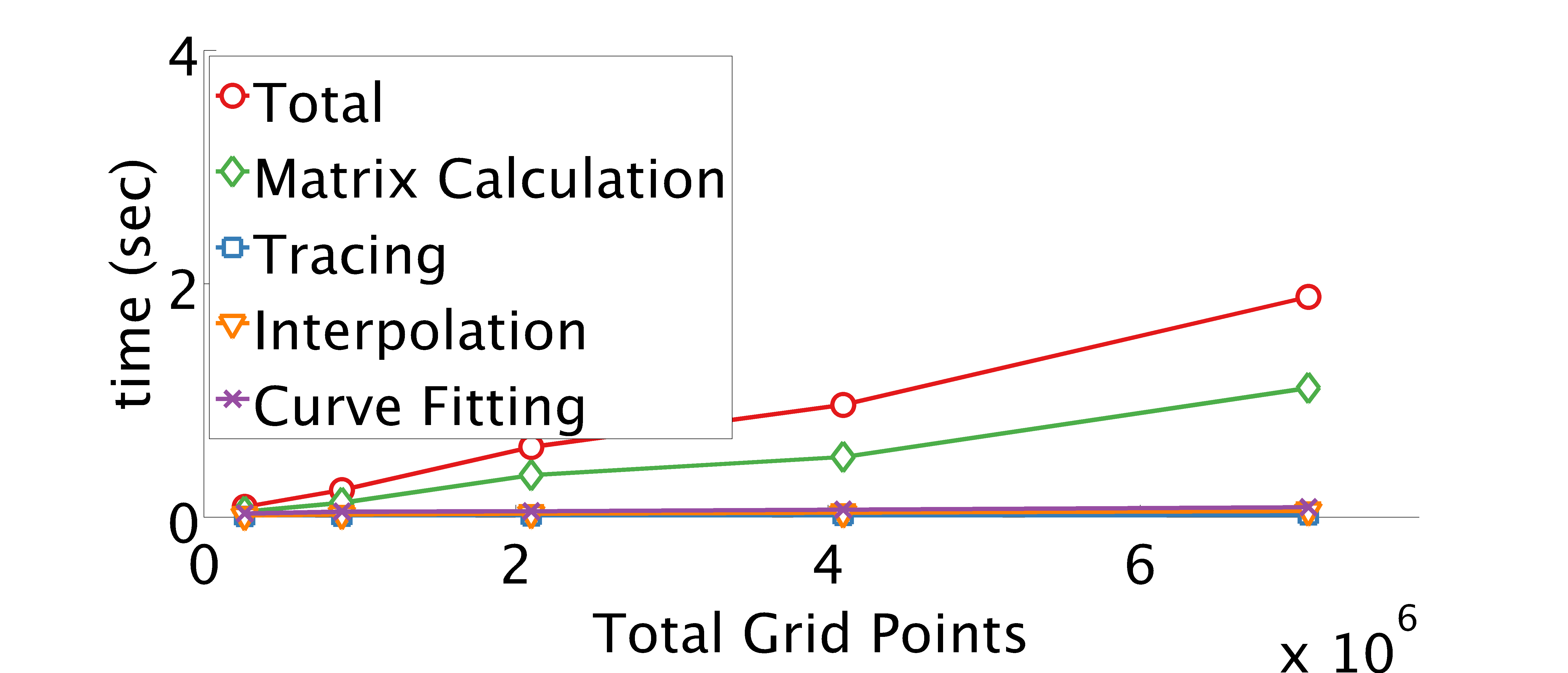}
    \caption{Calculation time as a function of increasing the number of grid points in the data set.}
    \label{fig:gridtest}
 \end{figure}

In Figure \ref{fig:gridtest}, we show how the algorithm scales with increasing data set size.    Grid point sizes of $64^3, 96^3, 128^3, 160^3$, and $192^3$ were tested.  Over all these data sets, the number of vortices was kept constant at two, while the data set size was increased.  In this dilute vortex state, with a small, fixed number of features to find, the bulk of the algorithm time is performing the matrix calculation.  Both calculations scale linearly with the number of grid points.  Thus the total time also scales linearly with the number of grid points, when the number of features is kept constant and is small.

\subsection{Scaling with Number of Vortices} \label{scalevortex}

\begin{figure}[htb]
        \centering
    \includegraphics[width=3.5in]{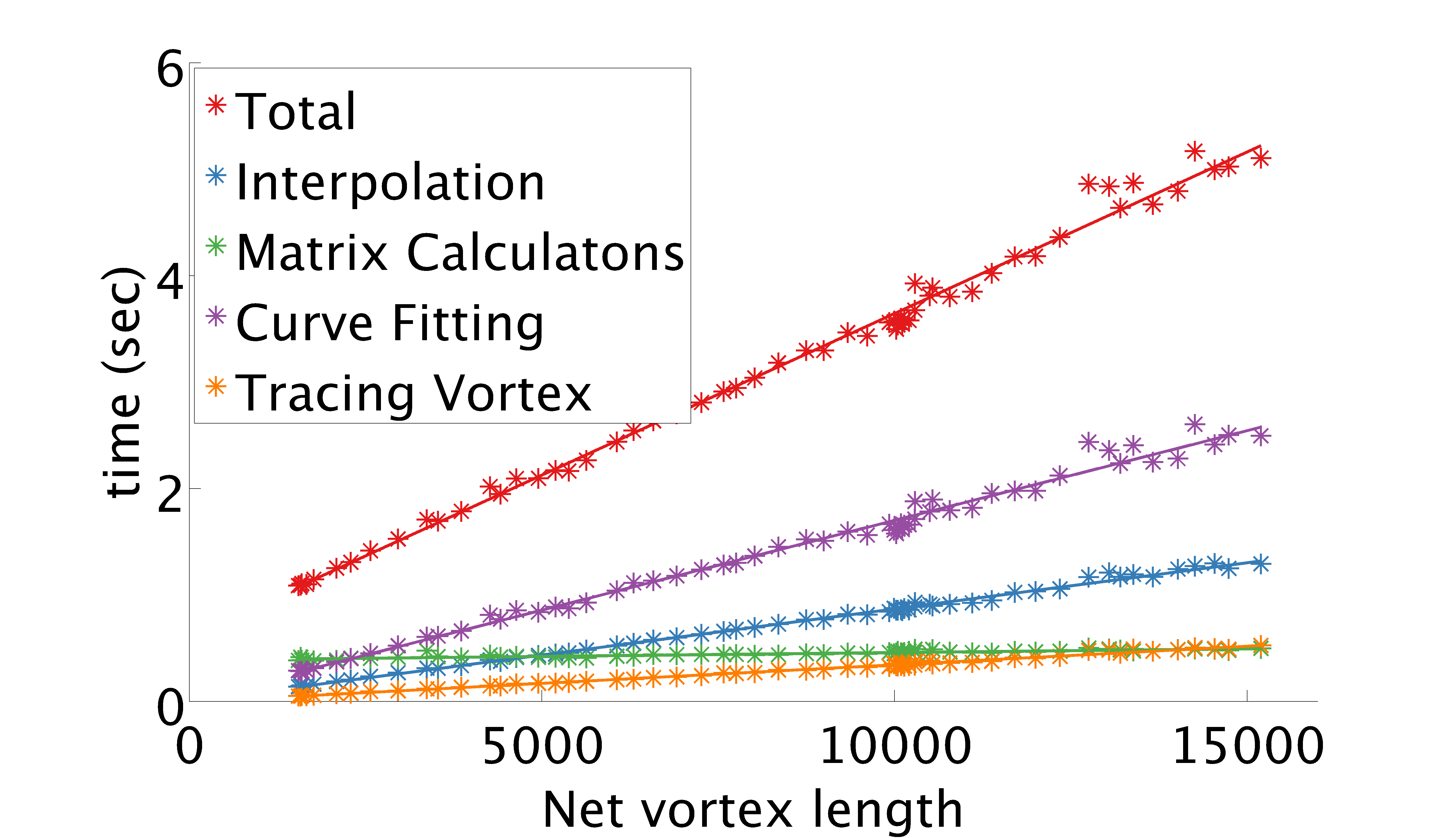}
    \caption{Scaling of parts of the algorithm as the vortex length increases for a fixed number of grid points.}
    \label{fig:vortextest}
 \end{figure}

The performance of steps 1-4 of the topological extraction method described above does not depend on the topology of vortices.  These steps are invariant to factors such as the direction or the tortuous path of a vortex.  They do, however, depend on the net vortex length in the data. The fifth step of the algorithm, in contrast, does depend on the topology of the vortices; this determines the number of recursion iteration steps required to fit the vortex.  However, here we focus primarily on the scaling of the algorithms relative to net vortex length. In Figure \ref{fig:vortextest}, the size of the mesh (128x128x128) was kept fixed while increasing the number of vortices present.  The triangulation interpolation method was used.  As can be seen, the matrix calculation is invariant to the increase in the number of features.  However, the time to trace the vortex structures, the time to calculate the interpolations, and the time to fit Bezier curves increase linearly with the net length of the vortices, measured in puncture points. Fitting cubic Bezier curves and generating the higher-precision vortex structure by interpolating the puncture points on the punctured faces constitute the bulk of the computational time for the data set of a dense vortex state.  Since, over this data, the vortex length is being increased by adding vortices of approximately constant length in puncture points, not by adding puncture points to each vortex, the computational cost of fitting a cubic Bezier curve is linear to the number of vortices in the system, and therefore to the number of puncture points. The computational cost of interpolation is always linearly proportional to the number of puncture points.  The choice of  interpolation method determines only the coefficient of the linear dependence.  Thus the four interpolation timings of Table \ref{table} should accurately predict how using different interpolations methods will scale the interpolation time.

 In general, as the data set size increases, if the planar density of vortices stays constant, the apparent length (measured in mesh elements) of all the vortices will grow in proportion to the total number of grid points.  Therefore, as simulations approach the macroscale, the calculation of the matrix and the tracing/interpolating of the vortex points will stay in roughly the same balance to each other, with the matrix calculation dominating in sparse vortex states and the interpolation calculation dominating in dense vortex states.  The cost of fitting piecewise cubic Bezier curves, however, will grow and may dominate the calculation.  Both curve fitting and the interpolation of puncture points are optional.  Many  analyses, such as tracking and event detection, do not require the extra precision in the determination of the puncture point.   Additionally the choice to fit curves to the points depends on whether further data compaction or data smoothing is desirable relative to the additional computational cost.

\subsection{Memory Usage}

In general, the minimal set of data structures to support this algorithm --- that is, the dictionary of interpolated points, the dictionary that holds the subgraph, and the final set of vortex objects --- scale with the number of puncture points, $N_p$.  In turn, the number of puncture points is proportional to the number of vortices $N_v$ in the data and to the discretization of a single edge $n_x$, that is $N_p \propto N_v*n_x$.  Thus the additional memory footprint of this algorithm above the original mesh data depends primarily on the density of vortices in the system but is moderate in size compared with the size of the mesh data, which is proportional to $n_x^3$.    As a rule, even for very dense vortex states, the additional memory requirements to support the data structures to generate the vortex objects are far less than 10\% of memory requirements for the original mesh data.  The final representation of vortices generally is less than 0.1\% of the original mesh data.  

In calculating the contour integrals, we can choose to precalculate certain arrays, for example, slices of the gauge transformation array that are used many times for computational efficiency.  In general, the precalculated arrays add more memory pressure than the data structures hold vortex objects.  Determining the best tradeoff between precalculating certain arrays versus recalculating values on the fly can be adapted as needed.

%
%
%
%
%

\section{Conclusion}\label{sec:conclusion}

In this paper, we have presented a method that can exactly extract the topological defect lines from a data set of complex scalars defined over a mesh.  In our application, the topological defects correspond to vortex lines in a TDGL simulation of a type-II superconductor.  Compared with prior methods, which generate isosurfaces, our method provides reliable subgrid resolution of vortex positions even when the vortices are densely packed.  The centers of vortices are detected by using the phase rather than magnitude of the complex scalar field.  Integrals are performed along gauge-transformed closed paths to find individual points along the core of a vortex. The real and imaginary parts of the field are then used to interpolate higher precision points.  The points are topologically ordered along a single vortex line by the construction and tracing of a subgraph generated from the underlying the mesh geometry.  Each vortex is then transformed to a compact and mesh-independent representation by fitting a piecewise cubic Bezier curve through the points.  The number of fitted curves is dependent on the tortuosity of the vortex, rather than the mesh resolution.  While implemented here on a regular structured mesh that is aligned along the Cartesian axes, this method can be easily generalized to an unstructured mesh composed of arbitrarily oriented polygonal faces.  

This analysis permits details of vortex interactions to be understood at a finer detail than was previously possible. (1) It allows vortices that are very close together to be disambiguated and the details of their interaction revealed.  In reference \cite{VitaliDraft}, this method was used to examine the before and after of two reconnecting vortices, revealing how the vortices mutually bent into an antiparallel configuration before swapping parts and rapidly repelling each other.  (2) It allows vortices to be visualized inside the interior of pinning defects modeled as suppressions of the $\psi$ parameter.  (3) It provides a reduced representation of individual vortices from which geometrical properties such as length, curvature, and angle of pinning defect penetration can be unambiguously measured.  (4) It provides the basis for tracking vortices over simulations, measuring their flow velocity and detecting reconnections and pinning events.  Thus, the macroscopic behavior of the vortices can be related to the measured properties of the simulation.  (5) Additionally, this provides a greatly reduced representation of the vortex state of a superconductor to be stored, compared with storing the entire state of $\psi$.  As TDGL simulations increase in size so as to model experimentally relevant mesoscale superconducting phenomena, it will be critical to be able to store and visualize reduced representations of the data, or the generation and storing of simulation data will quickly overwhelm computational effort.

\acknowledgements{We thank Alexei Koshlev and Hanqi Guo for useful discussions and thank H.G. for the method of efficiently solving the inverse bilinear interpolation.  We thank Sylvain Peyrefitte and Volker Poplawski for providing python implementations of the Ramer-Douglas-Peucker algorithm and Schneider algorithm, respectively.
This material was based upon work supported by the U.S. Department of Energy, Office of Science, Office of Advanced Scientific Computing Research, Scientific Discovery through Advanced Computing (SciDAC) program and the Materials Sciences and Engineering Division.
 C.L.P. was funded by the Office of the Director through the Named Postdoctoral Fellowship Program (Aneesur Rahman Postdoctoral Fellowship), Argonne National Laboratory.}\\

\noindent{\bf{Appendix A - Gauge-Invariant Vortex Detection}}\label{sec:gauge}

The total vorticity for $\psi=|\psi|e^{\imath\theta}$ is defined as
\begin{equation}
2\pi n\equiv -\oint_{\cal C} d{\bf l}\cdot\nabla\theta\,,\label{contourintegral}
\end{equation}
along a closed contour ${\cal C}$ with ${\cal C}=\partial{\cal A}$ (${\cal A}$ being the area enclosed by contour ${\cal C}$).

However, whereas the magnitude of $\psi$ is gauge invariant, the phase of $\psi$ is not.  In order to calculate the above line integral in a gauge-invariant manner, let us look at the expression
\[
{\bf j}_s\equiv {\bf J}_s/|\psi|^2\,.
\]
The supercurrent ${\bf J}_s$ is well defined and gauge invariant:
\[
{\bf j}_s=\frac{1}{2\imath|\psi|^2}\left(\psi^*\nabla\psi-\psi\nabla\psi^*\right)-{\bf A}=\nabla\theta-{\bf A}\,,
\]
and therefore
\begin{equation}
\oint_{\cal C} d{\bf l}\cdot\nabla\theta=\oint_{\cal C} d{\bf l}\cdot \left({\bf j}_s+{\bf A}\right) = \oint_{\cal C} d{\bf l}\cdot {\bf j}_s + \oint_{\cal C} d{\bf l}\cdot{\bf A}.
\end{equation}

The right-hand term then becomes $\Phi=\oint_{\cal C} d{\bf l}\cdot{\bf A}= \int \nabla \times {\bf A} \cdot d{\bf a}=  \int{\bf B}\cdot d{\bf a}$, or the total magnetic flux normal to the contour area.  So, now
\begin{equation}
2\pi n = -\oint_{\cal C} d{\bf l}\cdot {\bf j}_s -  \int{\bf B}\cdot d{\bf a}\, ,
\end{equation}
or the summation of two gauge-invariant integrals.

We use the expression
\[
{\bf j}_s=\frac{1}{|\tilde{\psi}|^2}{\rm Im}[{\tilde{\psi}}^*(\nabla - \imath{\bf A}){\tilde{\psi}}],
\]
where $\tilde{\psi}=\psi e^{\imath K x}$.   The phase factor, $e^{\imath K x}$, which makes $\tilde{\psi}$ a quasiperiodic function, is included in reference \cite{TDGL} so that the scalar potential $\mu$ does not have a discontinuity at the periodic boundary when a current is applied in the $x$-direction.  The value of $K$, which is time-varying but spatially invariant, is numerically calculated by the simulation and a provided quantity for this calculation.  We obtain
\begin{equation}\label{eq:js}
{\bf j}_s=\frac{1}{|\psi|}{\rm Im}[e^{-\imath(\theta+Kx)}(\nabla - \imath{\bf A}){|\psi|e^{\imath(\theta+Kx)}}]\,.
\end{equation}
Expanding this expresion, we write ${\bf j}_s={\rm Im}\left[e^{-\imath(\theta+Kx)}\nabla e^{\imath(\theta+Kx)}- \imath{\bf A}\right]=-{\bf A}+  K\hat{x} +\nabla\theta$, and thus
\begin{equation}
2\pi n= -\oint_{\cal C} d{\bf l}\cdot\left(\nabla\theta+  K\hat{x} -{\bf A}\right) - \int{\bf B}\cdot d{\bf a}.
\end{equation}

Another way to understand the derivation of this expression is to say that, since $\theta$ is dependent on the gauge, we choose a gauge and subsequent  value of $\theta$ along the contour to be the value where the non-current-induced part of the vector potential is zero.
The following is the expression for the gauge transformation for a Ginzburg-Landau system in the large $\lambda$ limit.
\begin{align} 
    \tilde{{\bf A}}({\bf r}) & = {\bf A}({\bf r}) + K\hat{x} - \nabla \chi \\
    \tilde{\mu}({\bf r}) & =  \mu({\bf r}) - x  \partial_t{K}
    \label{eq:mu_transformation} \\
    \tilde{\psi}({\bf r}) & = \psi({\bf r}) \, e^{\imath Kx-\imath \chi}.
    \label{eq:psi_transformation}
\end{align}
Per Eq.(\ref{eq:psi_transformation}), the transformation to $\theta$ is
\begin{equation}
\tilde{\theta} = \theta + Kx -\chi
\end{equation}
and
\begin{equation}
\nabla \tilde{\theta} = \nabla\theta + K\hat{x} -\nabla\chi \label{newtheta}
\end{equation}

If expression (\ref{newtheta}) is substituted into Eq.(\ref{contourintegral}), then an additional term is required to restore Eq.(\ref{contourintegral}) to gauge invariance.  (Note that integrating $K\hat{x}$ around a closed loop is always zero.)

\begin{equation}
2\pi n\equiv -\oint_{\cal C} d{\bf l}\cdot[\nabla\theta + K\hat{x} -\nabla\chi]   -\oint_{\cal C} d{\bf l}\cdot\nabla\chi
\end{equation}

We chose the gauge along the contour  ${\cal C}$, to be $\nabla \chi =  {\bf A}({\bf r})$. 

The final expression,
\begin{equation}
2\pi n\equiv -\oint_{\cal C} d{\bf l}\cdot[\nabla\theta + K\hat{x} -A(l)]   -\int{\bf B}\cdot d{\bf a}, \label{finalcontour}
\end{equation}
always calculates the change in $\theta$ around the contour with zero additional phase due to the choice of gauge.  This allows larger contours to be used without the calculation becoming invalid.  This also supports the minimal error in the interpolation of the puncture point.

The value of Eq.(\ref{finalcontour}) can be exactly calculated over a set of connected segments $\{l_i\}$ forming a closed path, where $\theta$ is $\theta_{i-1}$ and $\theta_{i}$ at the endpoints of segment $l_i$, as long as $\tilde{\theta}$ does not change by more than $\pi$ along any one segment, namely,

\begin{equation}
2\pi n\equiv -\sum_1^m \Delta\tilde{\theta}_{i,i-1} -\int{\bf B}\cdot d{\bf a}, \label{sumcontour}
\end{equation} 
where 
\begin{equation}
\Delta\tilde{\theta}_{i,i-1} = \textrm{mod}(\theta_i - \theta_{i-1}  + (K\hat{x} - A(l))\cdot l_i +\pi,2\pi)-\pi.
\end{equation}

The modulo operation above, which maps $\Delta\tilde{\theta}_{i,i-1}$ into the range $[-\pi,\pi]$, is necessary because the difference between two angles has a countably infinite number of values.  As long as $\tilde{\theta}$ has not changed by more than $\pi$, then the smallest value in magnitude is the correct one.  This is also why this is a condition for the correctness of the entire calculation.

In the large $\lambda$-limit Ginzburg-Landau solver described in reference \cite{TDGL}, the vector potential was defined as a linear function either in the $x$ and $z$ direction or in the $y$ and $z$ direction.    If the summation of Eq.(\ref{sumcontour}) is calculated as a four-point calculation around the edges of mesh element faces of a regular Cartesian mesh, then the set of all local transformations can be represented as two or three multidimensional arrays that hold the values of $\int d{\bf l}\cdot[K\hat{x} -A(l)]$ for $d{\bf l} = d{\bf x}, d{\bf y},$ or, $d{\bf z}$ for all the mesh element edges of the mesh.

For an $xz$ magnetic field and correspondingly defined vector potential, the first multidimensional array is the set of $z$-direction transformations, where each element is defined as 
\begin{equation} 
G_{z}(i,j,k) = -B_x\bar{y}(j)h_z,
\end{equation}
and the second multidimensional array is the set of $x$-direction transformations, where each element is defined as  
\begin{equation}
G_{x}(i,j,k) = B_z\bar{y}(j)h_x+Kh_x,
\end{equation} 
where $\bar{y}(j) = h_y(j -\frac{n_y}{2})$ if the y-direction is periodic and $\bar{y}(j) = h_y(j -\frac{n_y-1}{2})$ if it is not. The variables $h_x$, $h_y$  and $h_z$ are the edge lengths of the mesh elements.  There is no transformation along the $y$-direction edge.

We can also calculate the value of $\int d{\bf l}\cdot[K\hat{x} -A(l)]$ along an arbitrary vector as
\begin{align}
g({\bf r}_1,{\bf r}_2) =& \Delta xK + \Delta x\left(\frac{\bar{y}(j_1)+ \bar{y}(j_2)}{2}\right)B_z\\ \nonumber
 & - \Delta z\left(\frac{\bar{y}(j_1) + \bar{y}(j_2)}{2}\right)B_x,
\end{align}
where ${\bf r}_2-{\bf r}_1=(\Delta x,\Delta y,\Delta z)$ and ${\bf r}_1$ and ${\bf r}_2$ have $j$ indices of $j_1$ and $j_2$, respectively.   Using this form, we can create arbitrary polygonal contour paths in the mesh for calculating the vorticity.

For an $yz$ magnetic field and correspondingly defined vector potential, the first multidimensional array is the set of $z$-direction transformations, where each element is defined as
\begin{equation} 
G_{z}(i,j,k) = B_y\bar{x}(i)h_z,
\end{equation}
and the second multidimensional array is the set of $y$-direction transformations, where each element is defined as,  
\begin{equation}
G_{y}(i,j,k) = -B_z\bar{x}(i)h_y,
\end{equation} 
and the final multidimensional array is the constant $x$-direction transformation
\begin{equation}
G_{x}(i,j,k) = Kh_x,
\end{equation} 
where $\bar{x}(i) = h_x(i -\frac{n_x}{2})$ if the x-direction is periodic and $\bar{x}(i) = h_x(j -\frac{n_x-1}{2})$ if it is not.

Again, we can calculate the value of $\int d{\bf l}\cdot[K\hat{x} -A(l)]$ along an arbitrary vector as
\begin{align}
g({\bf r}_1,{\bf r}_2) =& \Delta xK - \Delta y\left(\frac{\bar{x}(i_1) + \bar{x}(i_2)}{2}\right)B_z \\ \nonumber
&+ \Delta z\left(\frac{\bar{x}(i_1) + \bar{x}(i_2)}{2}\right)B_y,
\end{align}
where ${\bf r}_1$ and ${\bf r}_2$ have $i$ indices of $i_1$ and $i_2$, respectively.  

\hfill \break
\noindent{\bf{Appendix B - Quasiperiodic Boundary conditions}} \label{sec:bc}

For the $xz$ plane homogeneous magnetic field, the $y$-direction (if specified periodic) is quasiperiodic.  This means there is a phase shift in $\psi$ across the $y$ boundary, whose magnitude is dependent on the $x$ and $z$ coordinates.  Hence, the following correction needs to be added to the calculation of $\Delta \tilde{\theta}$ for any segment that straddles the quasiperiodic boundary:
\begin{equation}
QP_y(x,z) = -L_yB_zx + L_yB_xz,
\end{equation}
where $x$ and $z$ are the coordinates where the quasiperiodic boundary is crossed in a positive y-direction.  For the $y$-directed edges of a Cartesian mesh, $x = h_xi$ and $z=h_zk$.

Similarly, for the $yz$ plane homogeneous magnetic field, the x-direction (if specified periodic) is quasiperiodic, and the analogous correction for any $x$-direction edge that straddles the periodic boundary is
\begin{equation}
QP_x(y,z) = L_xB_zy- L_xB_yz.
\end{equation}

\hfill \break
\noindent{\bf{Appendix C - Interpolation}}

Here we review multiple ways that the center of a vortex core can be interpolated from the value of $\psi$ at three or four points.  These methods use the set of values of $\psi$ defined at points along a contour to predict where inside the area enclosed by the contour $|\psi| = 0$, or both the real and imaginary parts of $\psi$ are zero.

Note that in order to get accurate and consistent results with contour integral calculation, one value of $\psi$ should be selected as a reference point, and the subsequent values of $\psi$ should have their phases corrected in the same manner as in the contour integral calculation.  For example, if the reference point is $\psi_0=|\psi_0|e^{\imath\theta_0}$, and next point is $\psi_1=|\psi_1|e^{\imath\theta_1}$ and if the gauge-invariant phase difference calculated between the reference point and the next point is $\Delta \tilde{\theta}$, then the value used for the next point should be $\psi_1'=|\psi_1|e^{\imath\theta_0+\Delta \tilde{\theta}}$.

{\bf a) Triangulation}
Given a set of three or more points that describes a polygonal contour path, each segment can be examined to see whether it contains a zero in either the real or imaginary part of $\psi$, based on a linear interpolation along each segment.  If exactly two zero-points  are found for the real and imaginary components, respectively, then the intersection between the pair of lines connecting the two pairs of points predicts the location of the puncture point.  If more or fewer than two zero-points are found for the real or imaginary components, then the sign changes around the points needs to be examined more closely to determine how to connect the points with lines, or a different interpolation method should be used.  

If the polygonal path is arbitrarily oriented in space such that the two lines are in a 3D space and not projected to a known plane, then the floating-point representation of the lines will be sufficient to prevent the two lines from properly intersecting.  The intersection should be determined numerically in a least-squares sense; we refer to this as a generalized triangulation.

{\bf b) Inverse Bilinear Interpolation}

A bilinear interpolation allows the value of a function at a point to be interpolated from the value of the function at four coplanar (but not collinear) points.  Thus, the point where $Re(\psi)=0$ and $Im(\psi)=0$ can be solved by inverting the bilinear interpolation. 
Assuming the calculation is performed in unit square coordinate system, then we seek $(x,y)$ such that
\begin{align}
b_1+b_2x + b_3y + b_4xy  &= 0\\
c_1+c_2x + c_3y + c_4xy  &= 0,
\end{align}
where $b_1 = Re(\psi(0,0))$, $b_2 = Re(\psi(1,0) - Re(\psi(0,0)$, $b_3 = Re(\psi(0,1)  - Re(\psi(0,0)$, and $b_4 = Re(\psi(0,0) -Re(\psi(1,0) - Re(\psi(0,1) + Re(\psi(1,1)$.  The $c$ coefficients are similarly defined for the imaginary part of $\psi$.
 Since a bilinear interpolation is a quadratic function, it is not, generally speaking, invertible.  However, the problem can be reformatted as finding the solution to a generalized eigenvector problem.
\begin{align}
Av = \lambda Bv,
\end{align}
where $y=\lambda$,

\begin{equation}
v = 
\begin{pmatrix}
x\\
1
 \end{pmatrix},
\end{equation}
\begin{equation}
A=- 
\begin{pmatrix}
b2 &  b1 \\
c2 & c1
 \end{pmatrix},
\end{equation}
and
\begin{equation}
B=
\begin{pmatrix}
b4 &  b3 \\
c4 & c3
 \end{pmatrix}.
\end{equation}
By determining the eigenvalues and associated eigenvectors of this equation, and choosing the $(x,y)$ pair both inside the bounds [0,1], the puncture point can be found.

{\bf d) Inverse Barycentric Interpolation}

To calculate the puncture point $r$ in a triangle arbitrarily oriented in space, we represent the point in barycentric coordinates in a 3D simplex, $(\lambda_1, \lambda_2, \lambda_3, 0)$.  The final coordinate $\lambda_4$ = 0, because we are constraining our point to one triangle of the surface of the tetrahedron.   Let $\psi_1, \psi_2, \psi_3$ represent the value of the complex order parameter on the three grid points of the triangle, each of which has coordinates $r_1, r_2, r_3$, where $r_i$ = $(x_i, y_i, z_i)$.   

As $|\psi|$=0 at the puncture point, both the real and imaginary part of $\psi$ must be zero at the point.  Also, by the definition of baryocentric coordinates, $\lambda_1 + \lambda_2 + \lambda_3 = 0$.  Hence we solve the following equation for $(\lambda_1, \lambda_2, \lambda_3)$.
\begin{equation}
\begin{pmatrix}
  Re(\psi_1) & Re(\psi_2) & Re(\psi_3) \\
  Im(\psi_1) & Im(\psi_2) & Im(\psi_3) \\
  1 & 1 & 1
 \end{pmatrix}
 \begin{pmatrix}
  \lambda_1 \\
  \lambda_2 \\
  \lambda_3
 \end{pmatrix}
 =
 \begin{pmatrix}
0 \\
0 \\
1
 \end{pmatrix}
\end{equation}

We convert the coordinates  ($\lambda_1, \lambda_2, \lambda_3$) to r by
\begin{equation}
r=T
 \begin{pmatrix}
  \lambda_1 \\
  \lambda_2 \\
 \end{pmatrix}
 + r_3,
\end{equation}

where $T$ is
\begin{equation}
T=
\begin{pmatrix}
  x_1-x_3 & x_2-x_3   \\
  y_1-y_3 & y_2-y_3  \\
  z_1 -z_3 & z_2 -z_3 
 \end{pmatrix}.
 \end{equation}


\bibliography{template}
The submitted manuscript has been created by UChicago Argonne, LLC, Operator of Argonne National Laboratory (``Argonne").  Argonne, a U.S. Department of Energy Office of Science laboratory, is operated under Contract No. DE-AC02-06CH11357.  The U.S. Government retains for itself, and others acting on its behalf, a paid-up nonexclusive, irrevocable worldwide license in said article to reproduce, prepare derivative works, distribute copies to the public, and perform publicly and display publicly, by or on behalf of the Government.
\end{document}